\documentclass{PoS}
\usepackage[latin1]{inputenc}
\usepackage{amsmath}
\usepackage{booktabs}

\newcommand{\npi}{\mbox{$N\pi$} }
\newcommand{\pref}[1]{(\ref{#1})}
\newcommand{\GA}{G_{\rm A}}
\newcommand{\GP}{G_{\rm P}}
\newcommand{\GPt}{\tilde{G}_{\rm P}}
\newcommand{\epsA}{\epsilon_{\rm A}}
\newcommand{\epsPt}{\epsilon_{\rm \tilde{P}}}
\newcommand{\epsP}{\epsilon_{\rm P}}

\title{$\npi$-excited state contamination in nucleon 3-point functions using ChPT}

\ShortTitle{$\npi$-state contamination  using ChPT}

\author{\speaker{Oliver B\"ar}\\
       Institut f\"ur Physik, 
       Humboldt Universit\"at zu Berlin\\
       Newtonstra{\ss}e 15,
	D-12489 Berlin,
	Germany \\
       E-mail: \email{obaer@physik.hu-berlin.de}}

\abstract{The $\npi$-state contribution to nucleon 3-pt functions involving the pseudoscalar density $P(x)$ and the time component $A_4(x)$ of the axial vector current are computed to LO in ChPT. In case of the latter the $\npi$ contribution is O($M_N$) enhanced compared to the single-nucleon ground state contribution. In addition, a relative sign in two terms of the $\npi$ contribution leads an almost linear dependence on the operator insertion time, as it is observed in lattice data. In case of the pseudoscalar density the $\npi$ contribution is strongly dependent on the momentum transfer, leading to a distortion of the pseudoscalar nucleon form factor. Finally, the $\npi$ state contamination in the form factors result in a violation of the generalized Goldberger-Treiman relation as observed in various lattice calculations.}

\FullConference{The 37th Annual International Symposium on Lattice Field Theory - LATTICE 2019\\
		 June 16-22 2019,
		Wuhan, China.}

\begin{document}

\section{Introduction}

While physical point simulations eliminate the need for a chiral extrapolation of lattice data, some problems become more severe with the pions as light as in Nature. The signal-to-noise problem \cite{Parisi:1983ae,Lepage:1989hd} typically becomes more pronounced the smaller the pion mass is, limiting the euclidean time separations in the correlation functions measured in lattice simulations. At the same time the excited-state contamination due to multi-particle states involving light pions gets larger since the gap to the single-particle ground state shrinks with smaller pion mass. 

In a series of papers \cite{Bar:2015zwa,Bar:2016uoj,Bar:2016jof} baryon chiral perturbation theory (ChPT) \cite{Becher:1999he} has been employed to estimate the two-particle nucleon-pion ($N\pi$) state contamination in lattice estimates of the nucleon mass, nucleon charges and pdf moments.\footnote{For reviews see Refs.\ \cite{Bar:2017kxh,Bar:2017gqh}.} The size of this particular excited-state contamination depends on the  quantity one is looking at. As shown at last year's lattice conference \cite{Bar:2018akl} the induced pseudoscalar form factor $\GPt$ of the nucleon is particularly prone to an $N\pi$-state contamination. ChPT predicts that the lattice plateau estimate for this form factor systematically underestimates the physical form factor. This bias is momentum transfer dependent, the smaller $Q^2$ the larger the underestimation. As a result the lattice data show a flatter $Q^2$ dependence than the pion-pole-dominance (PPD) model suggests. Using the ChPT result to correct data from the PACS collaboration \cite{Ishikawa:2018rew} good agreement with the PPD model and experimental data is achieved.

In the following new results are presented for the $\npi$  contamination in (i) the nucleon 3-point (pt) function involving the temporal component $A_4(x)$ of the axial vector current and (ii) lattice estimates of the pseudoscalar form factor $\GP$. In addition, with the results for the two axial form factors we explicitly check for the role the $\npi$ contamination plays in the violation of the generalized Goldberger-Treiman relation observed in various lattice calculations \cite{Bali:2018qus,Rajan:2017lxk,Jang:2018lup}. Technical details and a full account of the results are given in \cite{Bar:2018xyi,Bar:2019gfx}.

\section{\npi state contribution in nucleon axial and pseudoscalar form factors}

The isovector nucleon axial form factors are defined by the matrix element of the local isovector axial vector current between single nucleon states,
\begin{equation}\label{DefFF}
\langle N(\vec{p}',s')|A_{\mu}^a(0)|N(\vec{p},s)\rangle = \bar{u}(p',s')\left(\gamma_{\mu}\gamma_5 \GA(Q^2) - i \gamma_5\frac{Q_{\mu}}{2M_N}\GPt(Q^2)\right)\frac{\sigma^a}{2}u(p,s)\,.
\end{equation}
The nucleon momenta $\vec{p},\vec{p}'$ in the initial and final state imply the euclidean 4-momentum transfer $Q_{\mu}=(i E_{\vec{p}^{\,\prime}} - i E_{\vec{p}},\vec{q}),\, \vec{q} = \vec{p}^{\,\prime}-\vec{p}$. We follow the kinematic setup $\vec{p}^{\,\prime}=0$ that is often chosen in numerical simulations. $\GA(Q^2)$ and $\GPt(Q^2)$ on the right hand side of \pref{DefFF} refer to the axial and induced pseudoscalar form factors, respectively. Analogously, the matrix element of the local iso-vector pseudoscalar density $P^a(x)$ defines the pseudoscalar form factor,
\begin{equation}\label{DefpseudoscalarFF}
m_q \langle N(p',s')|P^a(0)|N(p,s)\rangle = m_q \GP(Q^2)\bar{u}(p',s')\gamma_5 \frac{\sigma^a}{2}u(p,s)\,.
\end{equation}
Lattice calculations of the form factors follow a standard procedure.
It is based on the calculation of the nucleon 2-pt function and the nucleon 3-pt functions involving the axial vector current and pseudoscalar density, and the ratio of these correlation functions,
\begin{eqnarray}\label{DefRatio}
R_{\mu}(\vec{q},t,t')& =&\frac{C_{3,X_{\mu}}(\vec{q},t,t')}{C_2(0,t)}\sqrt{\frac{C_2(\vec{q},t-t')}{C_2(0,t-t')}\frac{C_2(\vec{0},t)}{C_2(\vec{q},t)}\frac{C_2(\vec{0},t')}{C_2(\vec{q},t')}}\,,\quad \mu=1,\ldots 4,P\,.
\end{eqnarray}
$\mu=1,\ldots 4$ refers to the ratio with the axial vector current 3-pt function, and $\mu=P$ to the case with the pseudoscalar density. The euclidean times $t$ and $t'$ denote the source-sink separation and the operator insertion time, respectively.
The ratios are defined such that, in the asymptotic limit $t,t', t-t'\rightarrow \infty$, they converge to constant asymptotic values 
$R_{\mu}(\vec{q},t,t') \rightarrow\Pi_{{\mu}}(\vec{q})$. These are related to the form factors according to ($E_{N,\vec{q}}$: energy of a nucleon with momentum $\vec{q}$)
\begin{eqnarray}
\Pi_{{k}}(\vec{q})& =& \frac{i}{\sqrt{2E_{N,\vec{q}}(M_N+ E_{N,\vec{q}})}}\left( (M_N+E_{N,\vec{q}})\GA(Q^2) \delta_{3k}-\frac{\GPt(Q^2)}{2M_N} q_3q_k\right),\label{AsympValueR33}\\
\Pi_4(\vec{q})&=& \frac{q_3}{\sqrt{2E_{N,\vec{q}}(M_N+ E_{N,\vec{q}})}}\left(\GA(Q^2)+\frac{M_N-E_{N,\vec{q}}}{2M_N}\GPt(Q^2)\right)\,,\label{AsympValueR30}\\
\Pi_{\rm P}(\vec{q}) &=&  \frac{q_3}{\sqrt{2E_{N,\vec{q}}(M_N+ E_{N,\vec{q}})}}\,\GP(Q^2)\,.\label{AsympValueP}
\end{eqnarray}
The form factors are easily obtained from these expressions. In practice one only has access to the ratios $R_{\mu}(\vec{q},t,t')$ at time separations $t,t'$ far from being asymptotically large. In that case the correlation functions and the ratios contain a contribution from excited states with the same quantum numbers as the nucleon. Instead of the true form factors one is interested in one obtains {\em effective} form factors including an excited-state contamination,\footnote{For brevity we introduce the notation $G_{\tilde{\rm P}}=\tilde{G}_{\rm P}$.} 
\begin{eqnarray}
G^{\rm eff}_{\rm X}(Q^2,t,t')\, = \,G_{\rm X}(Q^2)\bigg[ 1 + \epsilon_{\rm X}(Q^2,t,t')\bigg],\quad X\,=\,A,P,\tilde{P}\,.\label{EffFF}
\end{eqnarray}
The excited-state contribution $\epsilon_{\rm X}(Q^2,t,t')$ vanishes for $t,t',t-t'\rightarrow \infty$. 

The dominant excited-state contamination for physical pion mass and large time separations is due to two-particle $\npi$ states. It can be computed in ChPT, the low-energy effective theory of QCD \cite{Tiburzi:2009zp}. Working at leading order (LO) in SU(2) baryon ChPT this theory contains a nucleon doublet with the proton and neutron fields and the three mass degenerate pions (we assume isospin symmetry). There is  a single interaction vertex which implies the well-known one-pion-exchange-potential between a pair of nucleons.  At LO only a few low-energy coefficients (LECs) enter: the axial charge and the pion decay constant as well as the pion and nucleon masses. The $\npi$ contribution to the 2-pt and 3-pt functions and the ratio $R_{\mu}$ are straightforwardly computed, yielding explicit expressions for the $\npi$ contamination $\epsilon_{X}(Q^2,t,t')$. Using the known phenomenological values for the LECs ChPT makes predictions for the $\npi$-state contamination present in lattice calculations of the form factors.

\section{Axial vector 3-pt function with temporal component $A_4$}

\begin{figure}[t]
\begin{center}
%\vspace{-0.3cm}
$R_4(\vec{q},t,t')$\\
\includegraphics[scale=0.4]{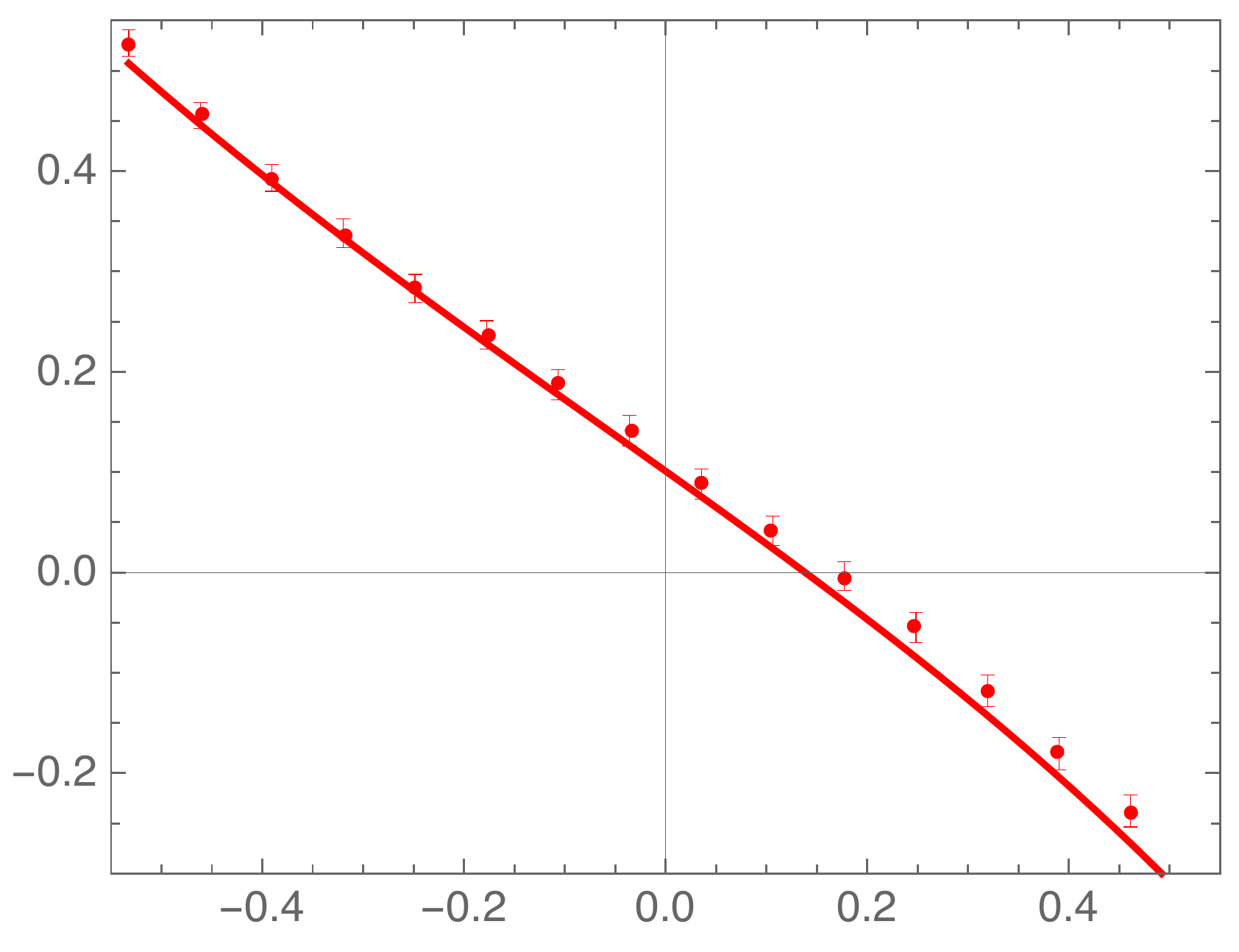}\\
$t'-t/2$ [{\rm fm}]
\caption{\label{fig:1} RQCD data \cite{Bali:2018qus} for the correlation function ratio $R_4(\vec{q},t,t')$ (red data points) for $t=1.06$ fm  and the ChPT result (red line) \cite{Bar:2018xyi}. }
\end{center}
\end{figure}

In principle lattice data for the 3-pt function involving the temporal component $A_4$ and the associated ratio $R_4$ may be used in the calculation of the axial form factors, see eq.\ \pref{AsympValueR30}. In practice, however, the data are usually excluded for large statistical errors and because of a large excited-state contamination. Examples for $R_4(\vec{q}, t,t')$ data can be found in \cite{Bali:2018qus,TschulzLat18}, and also in Y.-C.\ Jang's contribution to this conference \cite{YCJangLat19}. The ratio $R_4(\vec{q}, t,t')$ shows, for fixed $\vec{q}$ and $t$, a nearly linear dependence on the operator insertion time $t'$. In contrast to the other ratios there does not exist a plateau estimate. Thus, it is not obvious how to extract the single-nucleon (SN) state contribution.

In ChPT the 3-pt function is the sum of the SN and the $\npi$ contribution, 
\begin{equation}\label{C34}
C_{3,\mu=4}(\vec{q},t,t')  =  C^N_{3,\mu=4}(\vec{q},t,t') + C^{N\pi}_{3,\mu=4}(\vec{q},t,t')\,.
\end{equation}
If we perform the non-relativistic expansion for the nucleon energy, $E_{N,\vec{q}}\approx M_N +\vec{q}^2/2M_N$, we find
\begin{equation}
 C^N_{3,\mu=4} = {\rm O}\left({M_N^{-1}}\right)\,,\qquad  C^{N\pi}_{3,\mu=4} = {\rm O}(1)\,.
\end{equation}
Thus, the $\npi$ contribution in \pref{C34} is ``O($M_N$)-enhanced'' compared to the SN contribution, explaining the large excited-state contamination usually observed in lattice data for this correlation function. 
The origin of the large $\npi$ contribution is that the axial vector current can either directly create a pion at $t'$ that propagates to the sink where it is annihilated. Or a pion, created at the source, is destroyed by the axial vector current at $t'$.  It turns out \cite{Bar:2018xyi} that there is a relative sign between these two large contributions. As a consequence, $R_{4}(\vec{q},t,t')$ displays a 
$\sinh\left[E_{\pi,\vec{q}}\left(t'-{t}/{2}\right)\right]$ behavior, explaining the observed nearly linear dependence on $t'$.

Figure \ref{fig:1} shows the data of the RQCD collaboration \cite{Bali:2018qus} for $R_4(\vec{q}, t,t')$ (red data points) as a function of the (shifted) operator insertion time $t'-t/2$ for fixed $t=1.07$ fm and for $\vec{q}$ corresponding to  a momentum transfer $Q^2 =0.073$ GeV$^2$. 
The LO ChPT result for the ratio is shown by the red line. We emphasize that the ChPT result is not a fit to the lattice data, all LECs are fixed by their phenomenological values. 

Apparently, ChPT describes the data very well, in fact better than expected.
The time separations $t-t'$ and $t'$ need to be sufficiently large such that pion physics dominates the 3-pt function, and we can naively expect a minimal separation of about 1 fm for both $t-t'$ and $t'$. Therefore, for a source-sink separation as small as $t\approx 1$ fm we may expect ChPT to describe the 3-pt function and the ratio for $t'$ close to $t/2$, if at all.  The good agreement for all $t'$, even close to source and sink, is surprising and deserves further study.

\section{Pseudoscalar form factor}

\begin{figure}[t]
%\vspace{-0.3cm}
\hspace{2cm}$ \epsilon^{N\pi}_{X}(Q^2,t,t'),\, X=A,P,\tilde{P}$\hspace{5cm} $\GP^{\rm norm}(Q^2,Q_{\rm ref}^2,t)$\\
\includegraphics[scale=0.49]{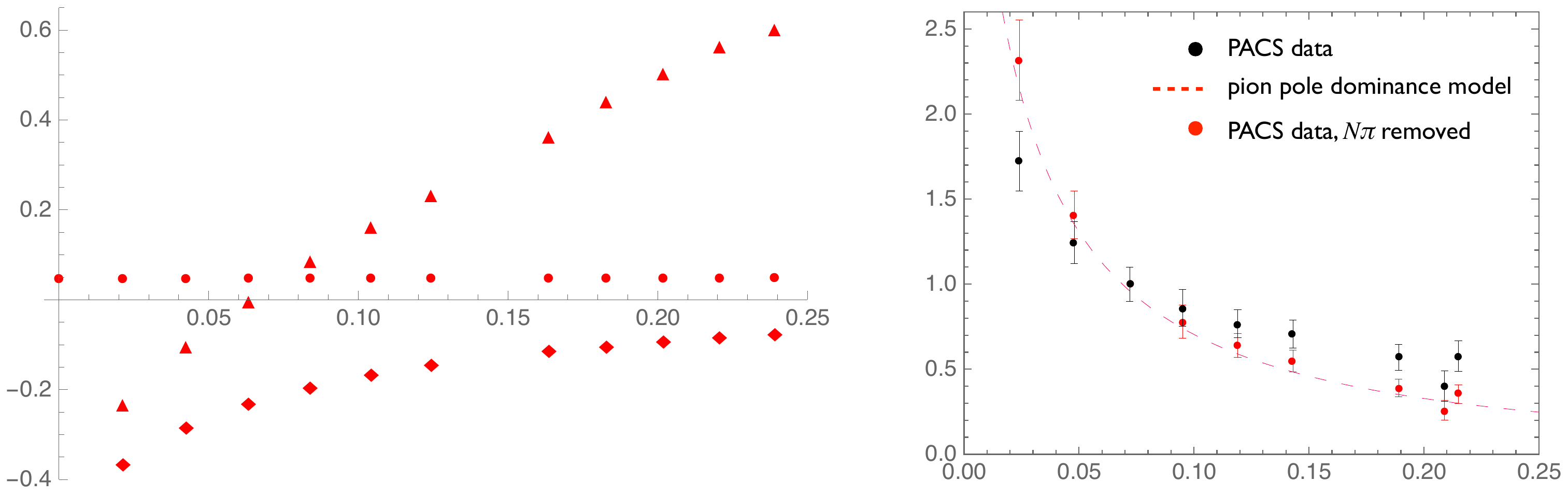}\\
\phantom{p}\hspace{3.5cm} $Q^2/{\rm GeV}^2$\hspace{6cm} $Q^2/{\rm GeV}^2$\\[-2ex]
\caption{\label{fig:2} Left panel: The relative deviations $\epsA(Q^2,t,t')$ (circles), $\epsPt(Q^2,t,t')$ (diamonds) and $\epsP(Q^2,t,t')$ (triangles) as a function of $Q^2$ for $t=2$ fm and $t'=t/2$. Discrete $Q^2$ values for finite spatial volumes satisfying $M_{\pi}L=6$.
Right panel: PACS data from \cite{Ishikawa:2018rew} for the normalized pseudoscalar form factor $\GP^{\rm norm}(Q^2,Q_{\rm ref}^2,t)$ (black) for $t=1.3$ fm and $Q^2_{\rm ref} = 0.072\,{\rm GeV}^2$.  Red symbols show the data with the $\npi$ contamination removed. The dashed red line corresponds to the PPD result.}
\end{figure}

As a measure for the $N\pi$-state contribution we introduce the relative deviation of the effective form factor from the true form factors,
\begin{eqnarray}\label{DefEpsilons}
\epsilon^{\rm N\pi}_{\rm X}(Q^2,t,t')\equiv \frac{G^{\rm eff}_{\rm P}(Q^2,t,t')}{G_{\rm P}(Q^2)} -1\,, \qquad X=A,\tilde{P},P.
\end{eqnarray}
The left panel in figure \ref{fig:2} shows $\epsilon^{\rm N\pi}_{\rm X}$ for all three form factors. The source-sink separation is chosen $t=2$ fm and $t'=t/2$, corresponding to the so-called mid-point estimate for the form factors. The discrete values for the momentum transfer correspond to the lowest possible values for spatial volumes with $M_{\pi}L=6$, as it is realized in the lattice simulations of the PACS collaboration \cite{Ishikawa:2018rew}, for example. The results for the axial form factors (dots and diamonds) have already been presented in \cite{Bar:2018akl}, new are the results for the pseudoscalar form factor (triangles). Figure \ref{fig:2} shows that the bias of the mid-point estimate $\GP^{\rm mid}$ is momentum dependent: $\GP^{\rm mid}$ underestimates for $Q^2 \lesssim 0.06$ GeV$^2$ (up to $\approx -20\%$ for the smallest momentum transfer), and overestimates for $Q^2\gtrsim0.06$ GeV$^2$  (up to $\approx +50\%$ for the largest momentum transfer in the figure). 

To compare  with lattice data it is useful to consider the normalized pseudoscalar form factor 
\begin{equation}\label{GPnorm}
\GP^{\rm norm}(Q^2,Q_{\rm ref}^2,t)\equiv \frac{\GP^{\rm plat}(Q^2,t)}{\GP^{\rm plat}(Q_{\rm ref}^2,t)}\,,
\end{equation}
which is independent of $Z_{\rm P}$. The right panel in figure \ref{fig:2} shows PACS lattice data \cite{} for this ratio (black symbols) for $Q^2_{\rm ref} = 0.072(2)\,{\rm GeV}^2$.  The red dashed line shows the PPD result for this ratio. Even though the statistical errors are large the momentum transfer dependence of the lattice data displays a flatter $Q^2$ dependence than the PPD model, just as ChPT predicts for the impact of the $\npi$ state contamination. With the ChPT result we can  analytically remove the anticipated LO $N\pi$-state contamination from the lattice data. The result is shown by the red data points in fig.\ \ref{fig:2}, which show much better agreement with the PPD model. The good agreement is again surprising, given that the lattice data is obtained with a fairly short source-sink separation $t\approx 1.3$ fm.

\section{Generalized Goldberger-Treiman relation}

\begin{figure}[t]
%\vspace{-0.3cm}
\hspace{3cm}$r_{\rm PCAC}(Q^2,t)$\hspace{5.5cm} $r_{\rm PCAC}(Q^2,t)$\\
\includegraphics[scale=0.46]{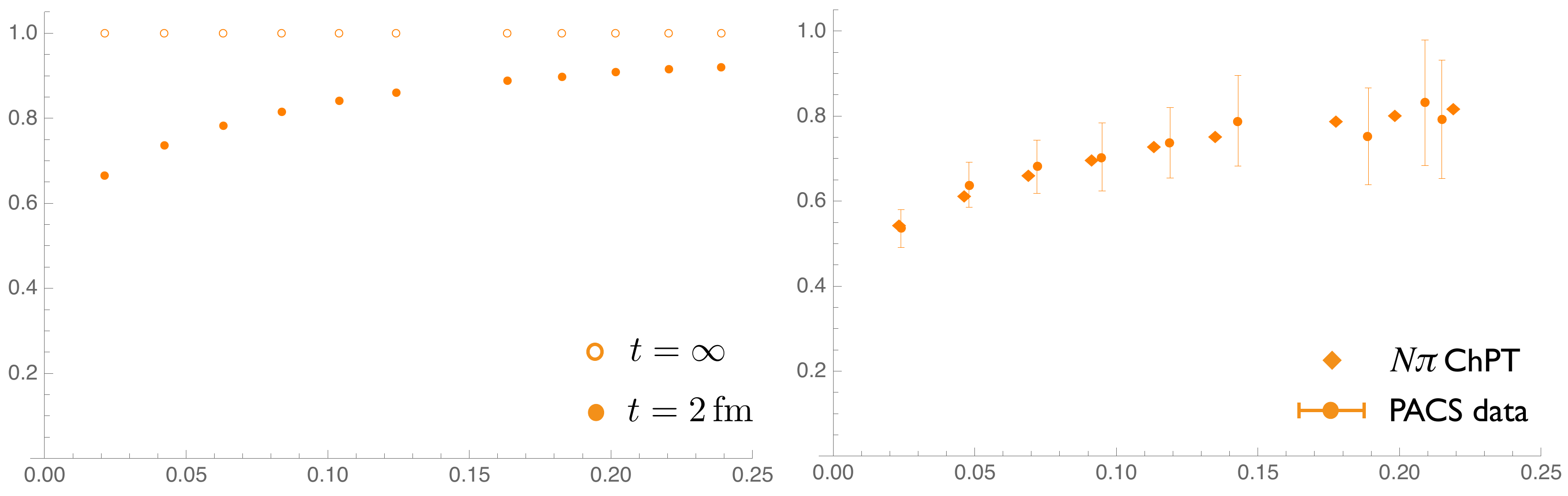}\\[-0.2ex]
\phantom{p}\hspace{3.5cm} $Q^2/{\rm GeV}^2$\hspace{6cm} $Q^2/{\rm GeV}^2$\\[-2ex]
\caption{\label{fig:3} Left panel: LO ChPT result for $r_{\rm PCAC}(Q^2,t)$ for $t=2$ fm (solid symbols) and $t=\infty$ (open symbols). 
Right panel: $r_{\rm PCAC}(Q^2,t)$, ChPT result (diamonds) compared to PACS data (dots).}
\end{figure}

The two axial and the pseudoscalar form factor are not independent. The partially conserved axial vector current (PCAC) relation, $\partial_{\mu}A_{\mu}^a(x) = 2m_q P^a(x)$, taken between SN states, leads to the so-called generalized Goldberger-Treiman (GGT) relation
\begin{equation}\label{pcacformfactorlevel}
2M_N G_{\rm A}(Q^2) -\frac{Q^2}{2M_N} \GPt(Q^2) = 2 m_q \GP(Q^2)
\end{equation}
between the three form factors.
This relation, however, is usually found to be violated by lattice estimates for the from factors. As a measure for this violation one may define the ratio
\begin{equation}
r_{\rm PCAC}(Q^2,t)=\frac{Q^2}{4M_N^2}\frac{\GPt^{\rm eff}(Q^2,t,t/2)}{\GA^{\rm eff}(Q^2,t,t/2)}+\frac{2m_q}{2M_N}\frac{\GP^{\rm eff}(Q^2,t,t/2)}{\GA^{\rm eff}(Q^2,t,t/2)}\,.
\end{equation}
If the effective form factors are equal to the physical ones $r_{\rm PCAC}$ is independent of the momentum transfer and equals 1. 
In practice, however, one typically finds a $Q^2$ dependent $r_{\rm PCAC} $ smaller than 1. The smaller the momentum transfer the larger the deviation from 1, see Refs.\ \cite{Bali:2018qus,Rajan:2017lxk,Jang:2018lup} and the plenary talk by T.\ Bhattacharya at this conference \cite{TBhattacharyaLat19}. 

Having the ChPT results for the $\npi$-state contamination in all three form factor estimates we compute $r_{\rm PCAC}$ and check for the role of $\npi$ states in the violation of the gGT relation. The left panel of fig.\ \ref{fig:3} shows the ChPT results for $t=2$ fm and $t=\infty$ (the discrete values for the momentum transfer are again the lowest ones associated with $M_{\pi}L=6$). We find that ChPT predicts $r_{\rm PCAC}(Q^2, t=2\,{\rm fm}) < 1$ with the deviation increasing the smaller $Q^2$ is, just as it is found in lattice simulations. The right panel of fig.\ \ref{fig:3} displays the ChPT result for smaller source-sink separation $t= 1.3$ fm, together with $r_{\rm PCAC}$ formed with the lattice results of the PACS collaboration in \cite{Ishikawa:2018rew} obtained at approximately this $t$. We observe good agreement with the data.  The dominant source for $r_{\rm PCAC}<1$  can be traced back to the large $\npi$ contamination in the induced pseudoscalar form factor and the underestimation of $\GPt(Q^2,t)$, see fig.\ \ref{fig:2}, left panel.

\section{Conclusions}
%ChPT predicts significant $\npi$-state contaminations in various nucleon form factor estimates at currently accessible source-sink separations. 
The ChPT results strongly suggest that many observed deviations between lattice estimates and experimental form factor results are due to the $\npi$-state contamination, for instance the violation of the generalized Goldberger-Treiman relation.

\end{document}